\documentclass[reprint, amsmath,amssymb, aps, superscriptaddress, prb]{revtex4-2}

\usepackage[english]{babel}
\usepackage{graphicx}
\usepackage{dcolumn}
\usepackage{dsfont}
\usepackage{bm}
\usepackage{xcolor}

\usepackage{hyperref}

\newcommand{\ZZ}{\mathrm{Z\kern-.3em\raise-0.5ex\hbox{Z}}}

\newcommand{\ket}[1]{\left| #1 \right\rangle}

\begin{document}
\title{Crystal field excitations in rattling clathrate CeBa$_7$Au$_6$Si$_{40}$.}

\author{Michelangelo Tagliavini}
\affiliation{Institute for Theoretical Physics, Heidelberg University, Philosophenweg 19, 69120 Heidelberg, Germany.}
\author{Federico Mazza}
\affiliation{Institute of Solid State Physics, Vienna University of Technology, Wiedner Hauptstr. 8-10, 1040 Vienna, Austria.}
\author{Xinlin Yan}
\affiliation{Institute of Solid State Physics, Vienna University of Technology, Wiedner Hauptstr. 8-10, 1040 Vienna, Austria.}
\author{Andrey Sidorenko}
\affiliation{Institute of Solid State Physics, Vienna University of Technology, Wiedner Hauptstr. 8-10, 1040 Vienna, Austria.}
\author{Kevin Ackermann}
\affiliation{Institute for Theoretical Physics, Heidelberg University, Philosophenweg 19, 69120 Heidelberg, Germany.}
\author{Andrey Prokofiev}
\affiliation{Institute of Solid State Physics, Vienna University of Technology, Wiedner Hauptstr. 8-10, 1040 Vienna, Austria.}
\author{Kurt Kummer}
\affiliation{European Synchrotron Radiation Facility (ESRF), F-380443, Grenoble, France.}
\author{Silke Paschen}
\affiliation{Institute of Solid State Physics, Vienna University of Technology, Wiedner Hauptstr. 8-10, 1040 Vienna, Austria.}
\author{Maurits W. Haverkort}
\affiliation{Institute for Theoretical Physics, Heidelberg University, Philosophenweg 19, 69120 Heidelberg, Germany.}

\date{\today}

\begin{abstract}
We investigate the local crystal-field environment of cerium in the clathrate compound CeBa$_7$Au$_6$Si$_{40}$ (Ce-BAS) using resonant inelastic x-ray scattering (RIXS) and magnetic susceptibility measurements. Ce-BAS is a rare example of a system where heavy-fermion physics coexists with low-energy rattling phonon modes, making it a candidate for enhanced thermoelectric performance. Magnetic susceptibility measurements reveal a temperature-dependent local moment that cannot be explained within a static crystal-field model. A fit to the susceptibility data requires strong mixing between the $j = 5/2$ and $j = 7/2$ crystal-field states, which implies large internal splittings inconsistent with RIXS spectra. In contrast, RIXS data are well described by a model with negligible $j = 5/2$–$j = 7/2$ mixing and an energy separation of 12~meV between the ground and first excited crystal-field states. The inability to reconcile these two datasets within a static framework points to a dynamical modification of the crystal-field potential. We attribute this to coupling between the Ce $4f$ electrons and low-energy phonons associated with the Ce rattling motion. This interpretation is consistent with theoretical predictions of phonon-enhanced Kondo effects and dynamical Jahn-Teller distortions. Our results highlight the need for multi-orbital impurity models that include phonon coupling to fully describe the low-energy physics of Ce-BAS.
\end{abstract}

\maketitle

\section{\label{sec:Intro}Introduction}

Ce-containing compounds exhibit a remarkable range of properties due to the interaction between local $4f$ states and extended band or ligand states. While most lanthanide ions are stable in a $3+$ valence, cerium can exist in either the Ce$^{3+}$ ($f^1$) or Ce$^{4+}$ ($f^0$) configuration. In the single impurity limit, the local moments of the Ce $f^1$ state act as scattering centers for band electrons, leading to the famous increase of the electrical resistivity with decreasing temperature first explained by J.\ Kondo \cite{kondo_resistance_1964, Hew97.1,burdin_coherence_2000, lawrence_two_1985, simeth_microscopic_2023}. For stoichiometric metallic crystals, the Kondo interaction leads to the formation of coherent bands with strongly reduced bandwidths near the Fermi energg, a key characteristic of heavy-fermion compounds \cite{goremychkin_coherent_2018,rahn_kondo_2022}. The resulting electronic and magnetic properties make Ce-based compounds attractive for various technological applications. For example, the narrow bands are associated with a sharply peaked density of states at the Fermi energy, which enhances the Seebeck \cite{Sch91.1,zlatic_theory_2005} and Nernst effect \cite{Beh16.1}. This makes Ce-based heavy-fermion compounds promising candidates for thermoelectric applications.

In clathrate materials, this thermoelectric advantage could in principle be combined with the suppression of lattice thermal conductivity via low-frequency ``rattling'' phonon modes of guest atoms in oversized host cages \cite{christensen_avoided_2008,Ike19.1}, with the prospect of achieving a significantly improved thermoelectric figure of merit. However, it has been a challenge to incorporate Ce into clathrate cages \cite{Kaw00.1,Pac01.1}, and only a dedicated single crystal growth technique allowed the successful synthesis of a Ce-containing clathrate, CeBa$_7$Au$_6$Si$_{40}$ (Ce-BAS) \cite{Pro13.1}. 

This compound can be thought of resulting from the parent compound Ba$_8$Au$_x$Si$_{46-x}$ (BAS), in which part of the Ba ions inside the Si cages are substituted with Ce. This substitution is also possible with La, giving rise to La-BAS, a non-magnetic reference compound of Ce-BAS. A significant enhancement of the Seebeck coefficient and figure of merit is observed in the Ce-substituted compound compared to the La-substituted compound, even at high temperatures \cite{Pro13.1}. Because an analysis of the specific heat gave a much lower Kondo temperature than consistent with this behavior, it was proposed that the effective Kondo temperature is enhanced by an interaction with the rattling mode \cite{Pro13.1}, an effect theoretically predicted by Hotta~\cite{hotta_enhanced_2007}.

In this work, we present a detailed analysis of the local crystal-field environment of Ce in Ce-BAS, as determined from resonant inelastic x-ray scattering (RIXS) and temperature-dependent magnetic susceptibility measurements. We find that the crystal-field parameters extracted from RIXS do not align with those inferred from the magnetic susceptibility. Specifically, the experimentally observed local magnetic moment is larger than predicted at low temperatures and smaller than expected at high temperatures. Although it is possible to fit the magnetic susceptibility data by invoking strong mixing between the $j = 5/2$ and $j = 7/2$ crystal field states, the resulting static crystal field parameters do not reproduce the RIXS data.

These observations support a scenario in which the Ce ion's motion within the Si cage,i.e., the rattling phonon mode, dynamically modifies the crystal-field potential and the associated low-energy electronic structure. This mechanism resembles the dynamic Jahn-Teller effect, where orbital degeneracies are lifted by coupling to lattice vibrations. Moreover, such dynamic phonon coupling may also enhance the Kondo interaction in nontrivial ways, as predicted by Hotta~\cite{hotta_enhanced_2007}, and should be included in order to gain a full theoretical understanding of Ce-substituted clathrate compounds.

\section{\label{sec:Symmetry}Symmetry and Local Moments}

CeBa$_7$Au$_6$Si$_{40}$ crystallizes in a cubic space group ($Pm\overline{3}n$). The Si and Au atoms form together two distinct cage-like substructures. The larger cage, centered at the $6c$ crystallographic site with point group $D_{2d}$ ($\overline{4}2m$), consists of 24 atoms and hosts a Ba atom at its center. The smaller cage, centered at the $2a$ site and associated with the $T_h$ ($m\overline{3}$) point group, forms a regular dodecahedron comprising 20 atoms arranged in 12 pentagons. Half of these dodecahedral cages are occupied by Ba atoms, while the other half are filled with Ce atoms~\cite{Pro13.1}.

In the $T_h$ point group, the Ce $4f$ orbitals split into three sets according to the irreducible representations of the cubic harmonics. The $f_{xyz}$ orbital transforms as the $a_u$ representation, while both the $f_{x^2}$, $f_{y^3}$, $f_{z^3}$ and the $f_{x(y^2-z^2)}$, $f_{y(z^2-x^2)}$, $f_{z(x^2-y^2)}$ orbitals transform as two distinct $t_u$ representations. Consequently, the local crystal field (i.e., the on-site energy structure of the $4f$ hybridized orbitals) can be parameterized by four quantities: the on-site energy of the $a_u$ state ($E_{au}$), the on-site energy of the first $t_u$ representation containing $f_{z^3}$ ($E_{tu_1}$), the on-site energy of the second $t_u$ representation containing $f_{x(y^2-z^2)}$ ($E_{tu_2}$), and a mixing parameter $M_{tu}$ describing the hybridization between the two $t_u$ representations. This mixing couples, for example, the $f_{z^3}$ orbital with $f_{x(y^2-z^2)}$, and likewise for the corresponding $f_{x^2}$, $f_{y^3}$ orbitals.

However, the cubic harmonics do not correspond to the true local eigenstates in the crystal. Spin-orbit coupling within the $4f$ shell ($\zeta = 76$~meV) introduces a splitting of approximately $7\zeta/2 \approx 270$~meV, which is significantly larger than the crystal-field splitting between the cubic harmonics, typically on the order of 10~meV in Ce compounds. Therefore, it is more appropriate to analyze the crystal field in the basis of $j$-$j_z$ coupled states rather than cubic harmonics.

In the $T_h$ point group, the $j = 5/2$ manifold splits into a $\Gamma_5^-$ doublet and a $\Gamma_6^-$ quartet. The corresponding crystal-field energies, expressed in terms of the parameters defined in the cubic harmonic basis, are given by:
\begin{align}
E_{\Gamma_5^-} &= \frac{3}{7} E_{au} + \frac{4}{7} E_{tu_2}, \\
E_{\Gamma_6^-} &= \frac{9}{14} E_{tu_1} + \frac{5}{14} E_{tu_2}.
\end{align}

The $j = 5/2$ crystal-field eigenstates in the $\Gamma$ basis are related to the $m_j$ states via:
\begin{align}
\renewcommand{\arraystretch}{1.6}
\label{eq:BasisRotation52}
\begin{pmatrix} 
\ket{\Gamma_5^-,1} \\ \ket{\Gamma_5^-,2} \\ 
\ket{\Gamma_6^-,1} \\ \ket{\Gamma_6^-,2} \\ 
\ket{\Gamma_6^-,3} \\ \ket{\Gamma_6^-,4} 
\end{pmatrix} = 
\begin{pmatrix} 
0 & \sqrt{\frac{5}{6}} & 0 & 0 & 0 & -\sqrt{\frac{1}{6}} \\
-\sqrt{\frac{1}{6}} & 0 & 0 & 0 & \sqrt{\frac{5}{6}} & 0 \\
\sqrt{\frac{5}{6}} & 0 & 0 & 0 & \sqrt{\frac{1}{6}} & 0 \\
0 & \sqrt{\frac{1}{6}} & 0 & 0 & 0 & \sqrt{\frac{5}{6}} \\
0 & 0 & 1 & 0 & 0 & 0 \\
0 & 0 & 0 & 1 & 0 & 0 
\end{pmatrix}
\begin{pmatrix} 
\ket{\frac{5}{2}, -\frac{5}{2}} \\ 
\ket{\frac{5}{2}, -\frac{3}{2}} \\ 
\ket{\frac{5}{2}, -\frac{1}{2}} \\ 
\ket{\frac{5}{2}, \phantom{-}\frac{1}{2}} \\ 
\ket{\frac{5}{2}, \phantom{-}\frac{3}{2}} \\ 
\ket{\frac{5}{2}, \phantom{-}\frac{5}{2}} 
\end{pmatrix}.
\end{align}

The $j = 7/2$ states in a $T_h$ point group split into two $\Gamma_5^-$ doublets and one $\Gamma_6^-$ quartet. In a $j$–$j_z$ basis, this decomposition introduces three additional parameters for the on-site energies and three mixing parameters for the $\Gamma_5^-$ doublets, as well as one mixing parameter for the two $\Gamma_6^-$ quartets—yielding a total of nine parameters. However, these parameters are not all independent: they are constrained by the four crystal-field parameters defined in the cubic harmonic basis—namely, the three distinct on-site energies and the mixing parameter between the two $t_u$ states.

The connection between the $j$–$j_z$ basis and the cubic harmonic basis can be established by performing a basis rotation of the crystal-field Hamiltonian. For the $j = 5/2$ manifold this rotation is given in equation \ref{eq:BasisRotation52}. For the $j = 7/2$ manifold, this rotation is given by:
\begin{widetext}
\begin{align}
\renewcommand{\arraystretch}{1.7}
\begin{pmatrix} 
\ket{2\Gamma_5^-,1} \\ \ket{2\Gamma_5^-,2} \\ 
\ket{3\Gamma_5^-,1} \\ \ket{3\Gamma_5^-,2} \\ 
\ket{2\Gamma_6^-,1} \\ \ket{2\Gamma_6^-,2} \\ 
\ket{2\Gamma_6^-,3} \\ \ket{2\Gamma_6^-,4} 
\end{pmatrix} = 
\begin{pmatrix} 
0 & 0 & -\sqrt{\frac{3}{12}} & 0 & 0 & 0 & \sqrt{\frac{9}{12}} & 0 \\
0 & -\sqrt{\frac{9}{12}} & 0 & 0 & 0 & \sqrt{\frac{3}{12}} & 0 & 0 \\
-\sqrt{\frac{5}{12}} & 0 & 0 & 0 & -\sqrt{\frac{7}{12}} & 0 & 0 & 0 \\
0 & 0 & 0 & \sqrt{\frac{7}{12}} & 0 & 0 & 0 & \sqrt{\frac{5}{12}} \\
0 & 0 & 0 & \sqrt{\frac{5}{12}} & 0 & 0 & 0 & -\sqrt{\frac{7}{12}} \\
\sqrt{\frac{7}{12}} & 0 & 0 & 0 & -\sqrt{\frac{5}{12}} & 0 & 0 & 0 \\
0 & \sqrt{\frac{3}{12}} & 0 & 0 & 0 & \sqrt{\frac{9}{12}} & 0 & 0 \\
0 & 0 & -\sqrt{\frac{9}{12}} & 0 & 0 & 0 & -\sqrt{\frac{3}{12}} & 0 
\end{pmatrix} 
\begin{pmatrix} 
\ket{\frac{7}{2}, -\frac{7}{2}} \\ 
\ket{\frac{7}{2}, -\frac{5}{2}} \\ 
\ket{\frac{7}{2}, -\frac{3}{2}} \\ 
\ket{\frac{7}{2}, -\frac{1}{2}} \\ 
\ket{\frac{7}{2}, \phantom{-}\frac{1}{2}} \\ 
\ket{\frac{7}{2}, \phantom{-}\frac{3}{2}} \\ 
\ket{\frac{7}{2}, \phantom{-}\frac{5}{2}} \\ 
\ket{\frac{7}{2}, \phantom{-}\frac{7}{2}} 
\end{pmatrix}.
\end{align}
\end{widetext}

However, these states are not the eigenstates of the $f^1$ crystal field Hamiltonian because multiple basis states belong to the same irreducible representation. Nevertheless, the Hamiltonian remains block-diagonal. For example, on the subspace of even or odd $\Gamma_6^-$ states—such as $\left(\ket{\Gamma_6^-,1}, \ket{\Gamma_6^-,3}, \ket{2\Gamma_6^-,1}, \ket{2\Gamma_6^-,3}\right)$ or $\left(\ket{\Gamma_6^-,2}, \ket{\Gamma_6^-,4}, \ket{2\Gamma_6^-,2}, \ket{2\Gamma_6^-,4}\right)$—the crystal-field Hamiltonian takes the form:
\begin{widetext}
\begin{align}
\label{eq:CH6}
H^{\Gamma_6^-} = 
\begin{pmatrix}
E_{\Gamma_6^-} -2\zeta & 0 & M_{t_u} & \frac{3\sqrt{5}}{4}(E_{2\Gamma_6^-} - E_{\Gamma_6^-}) \\
0 & E_{\Gamma_6^-} -2 \zeta & -\frac{3\sqrt{5}}{4}(E_{2\Gamma_6^-} - E_{\Gamma_6^-}) & M_{t_u} \\
M_{t_u} & -\frac{3\sqrt{5}}{4}(E_{2\Gamma_6^-} - E_{\Gamma_6^-}) & E_{2\Gamma_6^-} + \frac{3}{2} \zeta & 0 \\
\frac{3\sqrt{5}}{4}(E_{2\Gamma_6^-} - E_{\Gamma_6^-}) & M_{t_u} & 0 & E_{2\Gamma_6^-} + \frac{3}{2} \zeta
\end{pmatrix}.
\end{align}

Similarly, for the even or odd $\Gamma_5^-$ subspace—i.e., either $\left( \ket{\Gamma_5^-,1}, \ket{2\Gamma_5^-,1}, \ket{3\Gamma_5^-,1} \right)$ or $\left( \ket{\Gamma_5^-,2}, \ket{2\Gamma_5^-,2}, \ket{3\Gamma_5^-,2} \right)$—the Hamiltonian becomes:
\begin{align}
\label{eq:CH5}
H^{\Gamma_5^-} = 
\begin{pmatrix}
E_{\Gamma_5^-} -2\zeta & \frac{1}{2\sqrt{3}}(4E_{\Gamma_5^-} + 5E_{\Gamma_6^-} - 9E_{2\Gamma_6^-}) & \sqrt{\frac{4}{7}} M_{t_u} \\
\frac{1}{2\sqrt{3}}(4E_{\Gamma_5^-} + 5E_{\Gamma_6^-} - 9E_{2\Gamma_6^-}) & \frac{4}{3}E_{\Gamma_5^-} - \frac{3}{4}E_{2\Gamma_6^-} + \frac{5}{12}E_{\Gamma_6^-} + \frac{3}{2} \zeta & -\sqrt{\frac{3}{7}} M_{t_u} \\
\sqrt{\frac{4}{7}} M_{t_u} & -\sqrt{\frac{3}{7}} M_{t_u} & \frac{9}{4}E_{\Gamma_6^-} - \frac{5}{4}E_{2\Gamma_6^-} + \frac{3}{2} \zeta
\end{pmatrix}.
\end{align}
\end{widetext}

\section{\label{sec:CFParameters}Determining the Crystal-Field Parameters}

\begin{figure*}
    \centering
    \includegraphics[width=1\linewidth]{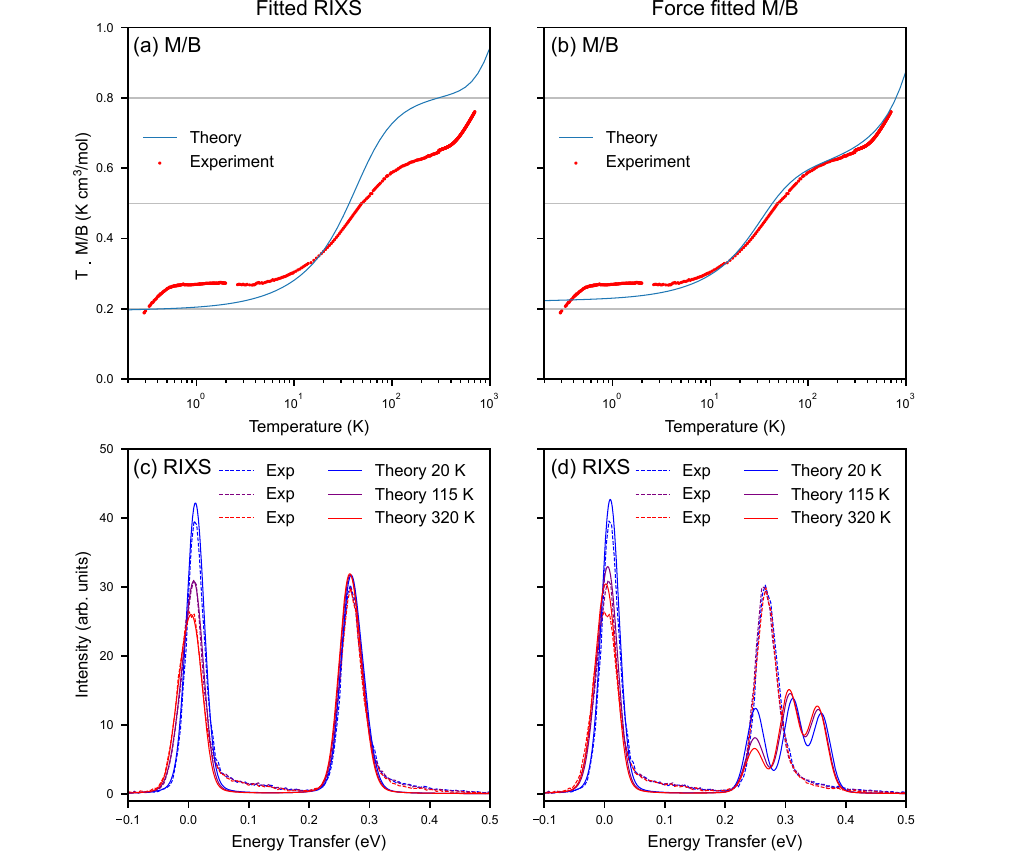}
\caption{
Comparison of experimental RIXS spectra (bottom panels) and magnetic susceptibility (top panels) with theoretical crystal-field calculations, using parameters either fitted to RIXS data (left column) or fitted to the magnetic susceptibility (right column). Panels (a) and (b): Experimental paramagnetic susceptibility of Ce-BAS (red), obtained by subtracting the susceptibility of La-BAS from that of Ce-BAS, shown as $T(M/B)$ for magnetic fields $B$ ranging from 0.05 to 9 Tesla. Theoretical predictions from crystal-field models are plotted in blue. Horizontal gray lines mark the expected local moments for a pure $j = 5/2$ $\Gamma_5^-$ doublet (0.2), $\Gamma_6^-$ quartet (0.5), and the full $j = 5/2$ manifold (0.8). Panels (c) and (d): Experimental RIXS spectra (dashed lines) and calculated spectra (solid lines) at 20~K, 115~K, and 320~K. In panel (c), the model is fitted to the RIXS data; in panel (d), the same parameters used for the susceptibility fit are applied. Discrepancies between theory and experiment in panel (d), particularly in the $j = 7/2$ excitation region, highlight the limitations of static crystal-field models when applied to Ce-BAS.
}

    \label{fig:RIXSChi}
\end{figure*}

Provided that the Ce $f$ orbitals give rise to essentially free moments, the crystal-field parameters can be determined through several experimental techniques. A relatively straightforward method is to measure the magnetic susceptibility which, in that case, reflects the temperature-dependent local magnetic moment. This moment varies depending on the crystal-field state. For a paramagnetic crystal, the temperature ($T$)-dependent magnetization is given by $M = (C/T) B$ for sufficiently small external magnetic fields $B$, with the Curie constant
\begin{align}
    C = \frac{n}{3 k_B} g_J^2 \mu_B^2 J(J + 1),
\end{align}
where $n$ is the number of atoms per unit volume, $k_B$ is the Boltzmann constant, $\mu_B$ is the Bohr magneton, and $g_J$ is the $g$-factor, which depends on both direction and the specific crystal-field state. Plotting $(M/B) \times T$ for a paramagnetic material yields a constant proportional to the crystal-field-dependent Curie constant. For an $f^1$ local moment with $j = 5/2$ in the $T_h$ point group, we expect Curie constants of approximately $C_{\Gamma_5^-} \approx 0.2$ K\,cm$^3$mol$^{-1}$, $C_{\Gamma_6^-} \approx 0.5$ K\,cm$^3$mol$^{-1}$, and $C_{j=5/2} \approx 0.8$ K\,cm$^3$mol$^{-1}$. At low temperatures, only the ground-state doublet or quartet is populated, yielding either 0.2 or 0.5 K\,cm$^3$mol$^{-1}$, while at higher temperatures, where both levels are thermally populated, the Curie constant approaches 0.8 K\,cm$^3$mol$^{-1}$.

In Fig.~\ref{fig:RIXSChi}, panels (a) and (b), we show the experimental magnetic susceptibility data, plotted as $M/B$ multiplied by $T$. The theoretically expected Curie constants discussed above are indicated by horizontal gray lines. The experiment, however, does not follow these expectations. Below 0.5 K, the system becomes non-magnetic, consistent with Kondo screening of the local moment. Between 0.5 and 10 K, a plateau is observed, corresponding to the effective moment of the ground-state multiplet. This value is close to, but somewhat larger than, that expected for a pure $\Gamma_5^-$ state. Between 10 and 100 K, the susceptibility increases, indicating a crystal-field excitation on the order of 10 meV. From 100 to 300 K, a second plateau appears in the $T M/B$ curve, corresponding to the effective moment of the $j = 5/2$ multiplet---albeit distinctly reduced compared to the theoretical full-multiplet value.

At the level of local crystal-field theory in $T_h$ symmetry, it is possible to modify the local moments through mixing between the $j = 5/2$ and $j = 7/2$ states. This mixing is controlled by the crystal-field parameter $M_{t_u}$. In Fig.~\ref{fig:RIXSChi}b (blue line), we show the result of fitting the magnetic susceptibility using such a model. The fitted crystal-field parameters are $E_{au} = 17$ meV, $E_{tu_1} = 45$ meV, $E_{tu_2} = 0$ meV, and $M_{t_u} = 80$ meV, resulting in a crystal-field splitting of 10 meV between the ground-state $\Gamma_5^-$ doublet and the first excited state. Since only relative energies are relevant, we can set one parameter—here, $E_{tu_2}$—to zero without loss of generality.

This model, with large $j = 5/2$–$j = 7/2$ mixing, allows a reasonable fit to the magnetic susceptibility. However, such a fit may not reflect the actual physical situation. Large mixing necessarily implies significant splitting within the $j = 7/2$ multiplet, as seen in Eqs.~\ref{eq:CH6} and~\ref{eq:CH5}, because $M_{t_u}$ directly couples the internal levels of the $j = 7/2$ states. While this scenario was proposed to be realized in CeRu$_4$Sn$_6$~\cite{amorese_determining_2018}, we will demonstrate in the next section---based on direct measurements of crystal---field excitations using RIXS—that Ce-BAS is not such a case.

To further investigate the magnitude of the crystal-field splitting, we performed RIXS experiments at beamline ID32 of the ESRF\cite{brookes_beamline_2018}. In RIXS, an x-ray photon is resonantly scattered from the sample. For Ce ions, this process involves exciting a $3d$ core electron into the $4f$ valence shell, promoting the Ce $f^1$ ground state into a $3d^9 4f^2$ intermediate configuration with several possible multiplet states. This excited state then decays either back to the ground state or into a low-energy excited final state~\cite{de_groot_resonant_2024}.

A central challenge in RIXS is the energy resolution. The excitation energy for promoting a $3d$ electron into the $4f$ shell is on the order of 900~eV, while the crystal-field excitations of interest correspond to an energy loss of only $\sim$10~meV. The RIXS measurements reported here were performed with a Gaussian energy resolution of $\sigma = 32$ meV—sufficient to directly resolve energy splittings of this order. Importantly, by rotating the sample during the experiment, we change the polarization of the incoming light relative to the crystal axes. This alters the selection rules and allows us to selectively enhance or suppress different $f$–$f$ transitions, thereby providing sensitivity to the symmetry and energy of the crystal-field excitations, even if we do not have the resolution to separately resolve each transition.

\begin{figure*}
    \centering
    \includegraphics[width=1\linewidth]{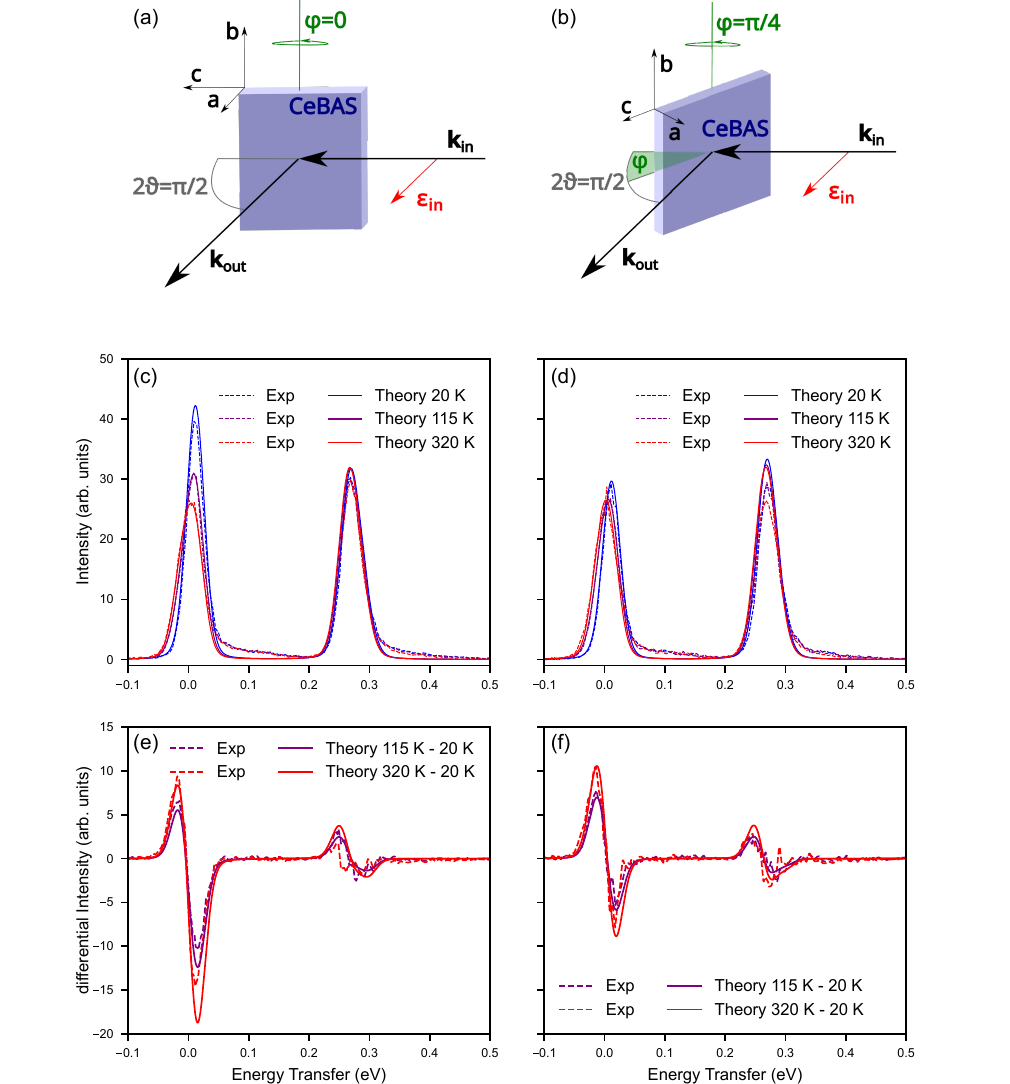}
 \caption{ Experimental RIXS spectra (dashed lines) measured at two different geometries: the geometry shown in panel (a) for the data in panels (c) and (e), and the geometry in panel (b) for panels (d) and (f). Panels (c) and (d) show temperature-dependent RIXS spectra measured at 20~K, 115~K, and 320~K for the two geometries. Panels (e) and (f) display the corresponding differential spectra, highlighting temperature-induced changes. All spectra are compared to theoretical calculations (solid lines) based on a crystal-field model, where the parameters $E_{au}$, $E_{tu_1}$, $E_{tu_2}$, and $M_{tu}$ are freely optimized as described in the text.
}

    \label{fig:RIXSFit}
\end{figure*}

In Fig.~\ref{fig:RIXSFit}, panels (a) and (b), we show the two geometries used in the RIXS experiment. The black arrows labeled $\mathbf{k}_{\mathrm{in}}$ and $\mathbf{k}_{\mathrm{out}}$ represent the wave vectors of the incoming and scattered light, respectively. The scattering angle was fixed at $90^\circ$, such that $\mathbf{k}_{\mathrm{out}}$ is parallel to the polarization vector of the incoming light, $\mathbf{\epsilon}_{\mathrm{in}}$. This configuration suppresses elastic scattering, which is primarily dominated by surface roughness. We used two polarization geometries: one with the incoming light polarized along a cubic crystallographic axis, and another rotated by $45^\circ$. Although the sample is cubic, these two geometries produce different spectra because RIXS probes a third-order and not a linear response function.

Panels (c) and (d) of Fig.~\ref{fig:RIXSFit} show the experimental spectra (dashed lines) for the two geometries at three different temperatures. Two prominent peaks are observed: one near zero energy loss and another near 270~meV. The first corresponds to $f$–$f$ transitions within the $j = 5/2$ manifold (i.e., between the $\Gamma_5^-$ doublet and $\Gamma_6^-$ quartet), while the second corresponds to transitions from the $j = 5/2$ to $j = 7/2$ states, which split into two $\Gamma_5^-$ doublets and one $\Gamma_6^-$ quartet. Although the crystal-field splitting is smaller than the experimental energy resolution (50~meV), we observe clear spectral changes with both temperature and light polarization (via sample rotation). For example, the intensity of the $5/2$ peak is significantly reduced in the geometry shown in the right column compared to that in the left column. These variations arise from state-dependent selection rules and enable us to extract crystal-field parameters by fitting both intensities and peak positions to theoretical calculations. Furthermore, with increasing temperature, the population of low-lying excited states gives rise to anti-Stokes peaks at negative energy loss. Although individual excitations are not fully resolved, a shift of the $5/2$ peak to lower energies with increasing temperature is clearly visible. These temperature-dependent changes become even more evident when examining the differential spectra shown in panels (e) and (f).

As a next step, we fitted the RIXS spectra to theoretical crystal-field calculations. This yielded the parameters $E_{au} = 6$~meV, $E_{tu_1} = 24$~meV, $E_{tu_2} = 0$~meV, and $M_{tu} = 15$~meV, resulting in a 12~meV energy splitting between the ground-state $\Gamma_5^-$ doublet and the first excited state, and only negligible mixing between the $j = 5/2$ and $j = 7/2$ manifolds.

We now return to the magnetic susceptibility shown in Fig.~\ref{fig:RIXSChi}. In panel (a), we plot the calculated susceptibility using the crystal-field parameters obtained from the RIXS fit. While the calculated energy splitting between the $\Gamma_5^-$ and $\Gamma_6^-$ states agrees well with experiment, the theoretical local moments deviate: the ground state moment is underestimated, and the high-temperature moment is overestimated. Conversely, in Fig.~\ref{fig:RIXSChi}d, we show RIXS spectra calculated using the parameters obtained from fitting the magnetic susceptibility. The excitation energy between the $\Gamma_5^-$ and $\Gamma_6^-$ levels was matched to the susceptibility step, yielding reasonable agreement for the low-energy $j = 5/2$ RIXS $f-f$ transition peaks near zero energy loss. However, the excitations into the $j = 7/2$ states are completely misrepresented. To achieve significant $5/2$–$7/2$ mixing with only three free parameters, one is forced to introduce artificially large splittings within the $j = 7/2$ multiplet. Thus, the scenario in which local moments are modified by strong static $j = 5/2$–$7/2$ mixing is clearly incompatible with the experimental RIXS data and can be definitively ruled out.

\section{\label{sec:Conclusion}Conclusion}

We have demonstrated that a static, local crystal-field model cannot simultaneously describe the crystal-field excitations observed in RIXS and the magnetic susceptibility data of Ce-BAS. When employing realistic crystal-field energies obtained from RIXS, the ground-state $\Gamma_5^-$ doublet exhibits a larger local moment in experiment than predicted by theory, whereas the high-temperature moment---where both the $\Gamma_5^-$ and $\Gamma_6^-$ states are thermally populated---is smaller in experiment than in theory.

Of course, deviations from local crystal field level schemes can occur in Kondo systems, where the magnetic moments cannot generally be considered as free. Their coupling to the band electrons leads to both indirect intermoment coupling (via the RKKY mechanism) and Kondo screening. The former will lead to deviations from simple Curie laws, the latter to both the screening of the local moment and to an enhanced Paul susceptibility. Whereas identifying the microscopic origin is beyond the scope of the present work, we would like to draw the attention to another effect that sets Ce-BAS apart from other Ce-based heavy fermion systems: the presence of low-energy ``rattling'' phonon modes~\cite{Pro13.1,Ike19.1}. Because in Ce-BAS, the Ce atoms reside within oversized cages, they undergo a rattling motion. Thus, it is intuitive that the $f$ moment of Ce will couple with the rattling mode. Theoretically, it was shown that the coupling between a magnetic impurity and a local Holstein phonon mode can dynamically enhance the Kondo interaction~\cite{hotta_enhanced_2007}. While Hotta’s model was limited to a single spin-orbital in the impurity model, similar physics is expected to be relevant in Ce-BAS due to the interplay between rattling phonons, conduction electrons, and local $4f$ moments.

Beyond modifying the effective Kondo temperature, phonon coupling is also expected to induce a dynamic contribution to the crystal field itself~\cite{haupricht_local_2010, zivkovic_dynamic_2024}. This effect can be interpreted as a competition between a dynamical Jahn-Teller distortion~\cite{slonczewski_theory_1963, ham_dynamical_1965} and the static crystal field that splits the $j = 5/2$ manifold into a $\Gamma_5^-$ doublet and $\Gamma_6^-$ quartet. To fully capture the underlying physics, it becomes essential to investigate a multi-orbital Anderson impurity model that is explicitly coupled to a local Holstein phonon mode.

This work is supported by the Deutsche Forschungsgemeinschaft (DFG, German Research Foundation) through the research unit QUAST, FOR 5249  P7 ID No. 449872909 and the Austrian Science Fund (FWF) through the projects I5868-N (FOR 5249 QUAST) and 10.55776/COE1 (quantA).


%

\end{document}